\documentstyle[12pt]{article}
\newcommand{\bm}[1]{\mbox{\boldmath$#1$}}
\title{On the grounding of spin effects \\
       in theory of synchrotron radiation }
\author{V.A.~Bordovitsyn\thanks{Corresponding author. E-mail:
bord@urania.tomsk.su}, A.N.~Myagkii\\ \it Tomsk State University,
634050 Tomsk, Russian Federation}
\date{}

\begin{document}
\begin{titlepage}
\maketitle
\begin{abstract}
The problem of the uniqueness in the introduction of spin
operators in the synchrotron radiation (SR) theory is discussed. For
this purpose we give the invariant spin projections on the basis of the
spin projections in the rest
frame. The spin equations are used to construct
the integrals of motion in the presence of the external electromagnetic
field.


\end{abstract}
\end{titlepage}

\section{Introduction}
\hspace*{\parindent} Synchrotron radiation (SR) is an unique instrument
in the research of individual properties of the relativistic
electrons~\cite{1}. In 1983 the first quantum spin-orientation-dependent
correction to the synchrotron radiation power was experimentally
identified at the Institute of Nuclear Physics, Siberian Branch of the
Academy of Sciences (Novosibirsk)~\cite{2,3}. In this experiment,
synchrotron radiation was for the first time observed to be dependent
on the spin orientation of a free electron moving in a macroscopic
magnetic field.  The procedure for experimental observation of spin
dependence of SR power was based on the Sokolov-Ternov radiative
self-polarization effect of the electrons.

To describe the spin effects in relativistic quantum theory one uses
spin operators.  The spin effects observed in the experiment are
determined by eigenvalues of the spin operators.  There are
different methods of introduction of the spin operators (see
Ref.~\cite{1}).  In this case the situation can arise when the same
projections of various spin operators lead to different physical
outcomes, such as time of the radiative self-polarization of the
relativistic electrons~\cite{4}. Hence, the procedure
of the unambiguity in the introduction of the spin operators has come into
importance.  This brings up the question: What spin operator is true?
This question is discussed in the paper on the basic of the invariant
correlations of the spin projection operators.
The paper is devoted to the memory of I.M. Ternov, who was
initiator of studies in this field.

\section{Spin operators of the Dirac particles}
\hspace*{\parindent}In Refs~\cite{5,6} it was shown that all
spin effects of SR can be obtained using wave function $\varphi$ of a free electron
with certain momentum and arbitrary spin orientation. That is why we
shall confine ourselves to the considering of spin operators of Dirac
particles with certain momentum $p^{\mu}$ or, in terms of
dimensionless momentum, $b^{\mu}=(\gamma,{\bm b})=p^{\mu}\left/ m_{0}c\right.$.

The simplest method of obtaining the spin operators is to separate from
the spin matrices $\gamma^{\mu}\gamma^{5}$ and $\sigma^{\mu\nu}$
\footnote{We use a metric
$g^{\mu\nu}={\rm diag}(-1,1,1,1)$ and 4-by-4 $\gamma$-matrices
$\gamma^{0}=i\rho_3, \gamma^{1}=i\rho_3\alpha_1,..., \gamma^{5}=-i\rho_1,
\sigma^{12}=\sigma_3, \sigma^{10}=-i\alpha_1,...,$ where $\rho_{i}$ and
${\bm\alpha}=\rho_{1}{\bm\sigma}$ are the well known Dirac matrices.}
the space-like parts (see in Ref.~\cite{8}). Then we obtain
\begin{eqnarray}
{\hat{\pi}}^{\mu}&=&( \gamma^{\mu}+i b^{\mu})\gamma^{5}=(
({\bm b}{\bm\sigma}),\rho_3{\bm\sigma}+\rho_1{\bm b} ), \nonumber\\
{\hat{\Pi}}^{\mu\nu}&=&\sigma^{\mu\nu}-(
\gamma^{\mu}b^{\nu}-\gamma^{\nu}b^{\mu} )=( \rho_3[ {\bm\sigma}{\bm b}
],{\bm\sigma}+\rho_2[ {\bm\sigma}{\bm b} ] ).
\label{2.4}
\end{eqnarray}
Spin operators satisfy commutative correlations for the generators of
little Lorentz group.

It can be shown that the mean values of the operators $\hat{\pi}^{\mu}$ and
$\hat{\Pi}^{\mu\nu}$ coincide with the classical four-vector of spin
$\pi^{\mu}=\left( \pi^{0},{\bm\pi} \right)$ and tensor of spin
$\Pi^{\mu\nu}=\left( {\bm\Phi},{\bm\Pi} \right)$.

The eigenvalues of the spin operators in the laboratory frame differ from
the eigenvalues $\zeta=\pm 1$ in the rest frame. For example,
\begin{equation}
\hat{\pi}_{z}\varphi=\zeta\,\sqrt{1+b^{2}_{z}}\,\varphi,\
\hat{\Pi}_{z}\varphi=\zeta\,\sqrt{1+b^{2}_{\bot}}\,\varphi,
\label{pzf}
\end{equation}
where $b^{2}_{\bot}=b^{2}_{x}+b^{2}_{y}$. In addition, in the laboratory
frame the operators $\hat{\pi}^{0}$ and $\hat{\bm\Phi}$ will induce
new eigenvalues
$\hat{\pi}^{0}\,\varphi=\zeta\,b\,\varphi$,
$\hat{\Phi}_{z}\,\varphi=\zeta\,b_{\bot}\varphi$
(see Ref.[9]), which vanish in the
laboratory frame. However, one can introduce such operator
$$\hat{\bm\Sigma}_{\sigma}={\bm\sigma}+(1-\rho_{3})\frac{i\left[
{\bm\alpha}{\bm b} \right]}{\gamma+1},$$
which in the laboratory frame will save the spin eigenvalues in the rest
frame, i.e.
\begin{equation}
({\bm\sigma}{\bm\nu})\varphi=\left(\hat{\bm\Sigma}_{\sigma}{\bm\nu} \right)\varphi=\zeta\,\varphi,
\label{snf}
\end{equation}
where ${\bm\nu}=(\sin\vartheta\cos\varphi,\sin\vartheta\sin\varphi,cos\vartheta)$
is the unit vector.

This operator is obtained as a result of unitary transformation of the
spin matrices ${\bm\sigma}$ (see Ref.~\cite{1}).
Similarly, one can
build the spin operator
$$\hat{\bm\Sigma}_{\rho_3{\sigma}}=\rho_3{\bm\sigma}+\rho_1\frac{\bm b}{\gamma}-
\frac{\rho_3{\bm b}({\bm\sigma}{\bm b})}{\gamma(\gamma+1)}.$$
It can be shown that spin operators $\hat{\bm\Sigma}$ possess
commutative properties of Pauli matrices~\footnote{One
can meet other operators possessing same properties (see in
Ref.~\cite{1}). These are Stech spin operator
$$\hat{\bm\Sigma}_{S}=\rho_3{\bm\sigma}+(1-\rho_3)\frac{{\bm b}({\bm\sigma}{\bm b})}{\gamma^{2}-1},$$
Foldy-Wouthuysen spin operator
$$\hat{\bm\Sigma}_{F-W}={\bm\sigma}-i\frac{\rho_3\left[ {\bm\alpha}{\bm b} \right]}{\gamma}-
\frac{\left[{\bm b}\left[ {\bm\sigma}{\bm b} \right] \
  \right]}{\gamma(\gamma+1)}$$
and so on. However, these operators coincide when applied to solutions of
the Dirac equation
$$\hat{\bm\Sigma}_{\sigma}\rightarrow\hat{\bm\Sigma}_{\rho_3{\sigma}}
\rightarrow
\hat{\bm\Sigma}_{S}\rightarrow\hat{\bm\Sigma}_{F-W}\rightarrow\hat{\bm\Sigma}.$$
}.

In order to transform the eigenvalues of $\hat{\bm\Sigma}$ into the
laboratory frame  the operator itself must be subjected to the Lorentz
transformations (see Ref.~\cite{1}).

\section{Invariant properties of spin operators}

\hspace*{\parindent} The spin operators
$\hat{\bm\Sigma}$,
$\hat{\Pi}^{\mu\nu}$,
and
$\hat{\pi}^{\mu}$ satisfy the invariant relations
\begin{equation}
\hat{\bm\Sigma}^2=\frac{1}{2}\,\hat{\Pi}_{\mu\nu}
\hat{\Pi}^{\mu\nu}=
\hat{\pi}_\mu\hat{\pi}^\mu=3.
\label{Sigma2}
\end{equation}

All the spin operators commutate with Hamiltonian of the free Dirac particle
with definite momentum $\hat\gamma=({\bm\alpha}{\bm b})+\rho_3.$
This allows to define completely a wave function
of particle with given momentum and concrete spin orientation.
But due to the noncommutativity of the spin operator
projections it is possible to measure simultaneously only the
eigenvalues for the absolute value and one of the spin
projections. It is known, however, (see, for example, in
Ref.~\cite{9}) that the same projection of the spin operators, for example
$\hat{\Sigma}_z$, $\hat{\pi}_z$ or $\hat{\Pi}_z$, gives the
various eigenvalues \footnote{For this reason some authors have
expressed some doubts concerning the correct application of
$\hat{\bf\pi}$ and $\hat{\bf\Pi}$ operators~\cite{10}.}. The
question arises: What projections of the spin operators
give the correct result? The answer to this question can be found by
constructing of relations between the projections of the spin operators.

As a spin projector one can take the antisymmetric tensor
$h^{\mu\nu}=(-{\bm e},{\bm h})$ with  components $h^{10}=-e_x,\dots$,
$h^{12}=h_z,\dots$\footnote{In the presence of the external
electromagnetic field this tensor can be identified with the tensor
$H^{\mu\nu}=(-{\bf E},{\bf H})$ (see, for example, in Ref.~\cite{11}).}.
In a general case $h^{\mu\nu}$ is not a space-like tensor, therefore the
space-like part should be separated out for further application
(here, $\rightarrow$ denotes the space-like part of a four-vector or a tensor)
$$\stackrel{\rightarrow}{h}\!{}^{\mu\nu}=\left(\gamma^2\left[{\bm\beta}
({\bm h}-[{\bm\beta}{\bm e}])\right], {\bm h}-\gamma^2\left[{\bm\beta}
({\bm e}+[{\bm\beta}{\bm h}])\right]\right)=
(-\stackrel{\rightarrow}{\bm e\vphantom{h}},\stackrel{\rightarrow}{\bm h}).$$

Applying the first nonzero invariant we come to the
dimensionless space-like tensor (for simplification we put ${\bm e}=0$)
\begin{equation}
S^{\mu\nu}=\,\stackrel{\rightarrow}{h}\!{}^{\mu\nu}\!\!\left/\!\sqrt{I_1}\right.=
\left([{\bm\beta}{\bm\eta}], {\bm\eta}-{\bm\beta}({\bm\beta}{\bm\eta})
\right)=({\bm Q},{\bm S}),
\label{Smunu}
\end{equation}
$$I_1=\frac{1}{2}\!\stackrel{\rightarrow}{h}\!{}_{\mu\nu}
\stackrel{\rightarrow}{h}\!{}^{\mu\nu}= \gamma^2[h^2-{(\bm\beta\bm h)}^2],
\quad {\bm\eta}=\frac{\gamma^2{\bm h}}{\sqrt{I_1}}=\frac{\gamma{\bm k}}
{\sqrt{1-\beta^2\cos^2\alpha}}, \quad ({\bm\beta}{\bm k})=
\beta\cos\alpha.$$

One can construct the
space-like four-vector corresponding to $S^{\mu\nu}$ \begin{equation}
s^\alpha=\frac{1}{2}\,\varepsilon^{\alpha\beta\mu\nu}S_{\mu\nu}b_\beta,
\quad S^{\mu\nu}=\varepsilon^{\mu\nu\alpha\beta}s_\alpha b_\beta.
\label{salpha}
\end{equation}

The components of $s^\alpha$ are equal to
$s^\alpha=\left(({\bm\beta}{\bm\eta}), {\bm\eta}\right)\left/\gamma \right.=(s^{0},{\bm s})$.
According to Eq.~(\ref{salpha}) they are associated with $S^{\mu\nu}$
by the relations
\begin{equation}
{\bm s}=\frac{1}{\gamma}\,{\bm S}+\gamma{\bm\beta}({\bm\beta}{\bm S}), \quad
{\bm S}=\gamma[{\bm s}-{\bm\beta}({\bm\beta}{\bm s})].
\label{vecs}
\end{equation}

In the rest frame the spin projectors ${\bm s}$ and ${\bm S}$
transform into the unit vector
$$
{\bm\nu}=\frac{1}{\gamma}\left[{\bm\eta}-\frac{\gamma}{\gamma+1}\,{\bm\beta}
({\bm\beta}{\bm\eta})\right]
$$
associated with ${\bm s}$
and ${\bm S}$ by Lorentz transformation.

One can show that the spin operators
$\hat{\Pi}{}^{\mu\nu}$ and
$\hat{\pi}{}^{\mu}$ satisfy the relations similar to
Eqs~(\ref{salpha}) and (\ref{vecs}).
On this basis one can obtain the following invariant relations
\begin{equation}
s_\alpha\hat{\pi}{}^\alpha=\frac{1}{2}\,S_{\mu\nu}\hat{\Pi}^{\mu\nu}=inv,
\label{sapa}
\end{equation}
\begin{equation}
\left({\bm\nu}\hat{\bm\Sigma}\right)=\frac{1}{\gamma}\,
\left({\bm S}\hat{\bm\pi}\right)=\frac{1}{\gamma}\,
\left({\bm s}\hat{\bm\Pi}\right).
\label{nuSigma}
\end{equation}

Note that the spin operators $\hat{\bm\Sigma}$,
$\hat{\pi}{}^\mu$, $\hat{\Pi}{}^{\mu\nu}$ in these relations can be
replaced by their classical analogue ${\bm\zeta}$, ${\pi}^\mu$,
${\Pi}^{\mu\nu}$.

We consider two extreme cases: 1)
${\bm\beta}\parallel{\bm\eta}\parallel{\bm
k}$, when ${\bm\beta}\parallel Z$ and $\beta_\perp=0$, 2)
${\bm\beta}\perp{\bm\eta}\parallel{\bm k}$, when ${\bm\beta}\perp Z$ and
$\beta_z=0$. In both cases ${\bm\nu}={\bm k}$ (spin is directed along
the $Z$-axis in the rest frame), however, the spin components are
different in the laboratory frame
\begin{eqnarray*}
\left({\bm k}\hat{\bm\Sigma}\right)&=&\frac{1}{\gamma}\,
\left({\bm k}\hat{\bm\pi}\right)=
\left({\bm k}\hat{\bm\Pi}\right),
\,\,\mbox{for}\,\, {\bm\beta}\parallel{\bm k},\\
\left({\bm k}\hat{\bm\Sigma}\right)&=&
\left({\bm k}\hat{\bm\pi}\right)=\frac{1}{\gamma}\,
\left({\bm k}\hat{\bm\Pi}\right),
\,\,\mbox{for}\,\, {\bm\beta}\perp{\bm k}.
\end{eqnarray*}
In terms of the eigenvalues of the spin operators these relations
coincide with $\zeta$.

In a similar way one can consider the kinematic spin
invariants which are induced by the particle motion
$$({\bm\sigma}{\bm\beta})=
\left(\hat{\bm\Sigma}{\bm\beta}\right)=\frac{1}{\gamma}\,
\left({\bm\beta}\hat{\bm\pi}\right)=
\left({\bm\beta}\hat{\bm\Pi}\right),$$
$$\left({\bm\nu}\hat{\bm\Phi}\right)=
\left({\bm s}\hat{\bm\Phi}\right)=\frac{1}{\gamma}\,
\left({\bm S}\hat{\bm\Pi}\right)=\frac{1}{\gamma}\,
\left[{\bm\eta}\hat{\bm\Phi}\right].$$

The first invariants characterize the spin projection onto the direction of
motion and the second invariants do the spin projection onto the direction of
$[{\bm k}{\bm\beta}]$.

One can show that in the BMT approximation (see Ref.~\cite{1}) the spin invariant of
Eq.(\ref{sapa}) remains valid in the presence of the external
electromagnetic field also (see Ref.~\cite{14}). In particular, in the
constant and uniform magnetic field when ${\bm h}\parallel{\bm
H}\parallel{\bm k}$, the spin projection onto the orbital radius is
characterized by the spin operator $\hat{\Phi}_z$.

\section{Spin invariants as integrals of motion}

\hspace*{\parindent}In the presence of the external electromagnetic
field the spin operators, generally speaking, stop being integrals of
motion.  It is known, for instance, that in the constant and uniform
electromagnetic fields they satisfy the operator equations similar to
the BMT equations~\cite{12}. Only some components of spin operators
remain the integrals of motion (see also Ref.~\cite{13}).  Nevertheless
one can show~\cite{14} that the spin invariants of Eq.~(\ref{sapa})
remain the integrals of motion if we choose the external
electromagnetic field tensor $H^{\mu\nu}$ as a projector $h^{\mu\nu}$.
Moreover, the spin invariants allow also to find such direction, for
which the spin projection of the moving particle will be always give
the same eigenvalue (the precession axis). Such a problem for electron
moving in the magnetic field of arbitrary configuration was for the
first time solved in Ref.~\cite{15}, where it was shown that the value of
$({\bm\nu}{\bm\Sigma})=const$, if the unit vector
${\bm\nu}$ (${\bm\beta}$, ${\bm H}$) satisfies the equation of BMT type. It
is a rather obvious result, as the vector ${\bm\nu}$ rotates with
respect to azimuth with the frequency of the spin precession.  We show
that this problem can be solved for projectors $s^\mu$ and $S^{\mu\nu}$
too. As an example we take the tensor $S^{\mu\nu}$. At the beginning we
represent the BMT equation in the form~\cite{1}
$$\frac{d\Pi^{\mu\nu}}{d\tau}=\frac{eg}{2m_0c}\,\widetilde{H}{}^{[\mu\rho}
\Pi_\rho^{\ \nu]},$$
where
$$\widetilde{H}{}^{\mu\rho}=H^{\mu\rho}+\frac{1}{c^2}\,v^{[\mu}v_\alpha
H^{\alpha\rho]}+\frac{2m_0}{egc}v^{[\mu}w^{\rho]}$$
is an effective electromagnetic field, $v^\mu$ and $w^\mu$ are the
four-velocity and acceleration, $\tau$ is the proper time, the square
brackets mean the antisymmetrization with respect to the corresponding
indexes. In the case of constant and uniform fields
$$\frac{d\widetilde{H}{}^{\mu\nu}}{d\tau}=\frac{g-2}{gc^2}\,
H^{[\mu\alpha}w_\alpha v^{\nu]}\neq 0,$$
but
$$\frac{d}{d\tau}\left(\widetilde{H}_{\mu\nu}\Pi^{\mu\nu}\right)=0,$$
that is rather obvious as
$\widetilde{H}_{\mu\nu}\Pi^{\mu\nu}=H_{\mu\nu}\Pi^{\mu\nu}$.
Taking into account
$S^{\mu\nu}\!=\,\stackrel{\rightarrow}{\widetilde{H}}\!\!{}^{\mu\nu}\!\!\left/
\!\sqrt{I_1}\right.$
one can find the such moving axes onto which the spin projection
will be constant in time.

\end{document}